\documentclass[twocolumn,aps,prb,showpacs,multicol,amssymb]{revtex4}

\usepackage[dvips]{graphicx}
\newcommand{\be}{\begin{equation}}
\newcommand{\ee}{\end{equation}}

\newcommand{\bea}{\begin{eqnarray}}
\newcommand{\eea}{\end{eqnarray}}
\newcommand{\bd}{\begin{displaymath}}
\newcommand{\ed}{\end{displaymath}}
\newcommand{\ba}{\begin{array}}
\newcommand{\ea}{\end{array}}
\newcommand{\bi}{\begin{itemize}}
\newcommand{\ei}{\end{itemize}}
\newcommand{\bc}{\begin{center}}
\newcommand{\ec}{\end{center}}
\newcommand{\bfl}{\begin{flushleft}}
\newcommand{\efl}{\end{flushleft}}
\newcommand{\bfr}{\begin{flushright}}
\newcommand{\efr}{\end{flushright}}

\def\6{\partial}  
   \def\e{\epsilon}

\def\m{\mu}

\def\no{\nonumber\\}
\def\bra{\langle}
\def\ket{\rangle}
\def\={\!\!\!&=&\!\!\!}
\def\+{\!\!\!&&\!\!\!+~}
\def\-{\!\!\!&&\!\!\!-~}

\begin{document}
\date{October  10, 2007}

\title{Dielectric Susceptibility and Heat Capacity of Ultra-Cold Glasses in Magnetic
Field}

\author
{Alireza Akbari$^{1,2}$, D. Bodea$^{2,3}$, and  A. Langari$^{4}$}

\affiliation{$^1$Institute for Advanced Studies in Basic Sciences,
P.O.Box 45195-1159, Zanjan, Iran
\\$^2$Max Planck Institute for the
Physics of Complex Systems, 01187 Dresden, Germany
\\$^3$Department of Physics,
``Babe\c{s}-Bolyai" University, 40084 Cluj Napoca, Romania
\\$^4$Department of Physics,
Sharif University of Technology, 14588-89694, Tehran, Iran}

\begin{abstract}
Recent experiments demonstrated unexpected, even intriguing
properties of certain glassy materials in magnetic field at low
temperatures. We have studied the magnetic field dependence of the
static dielectric susceptibility and the heat capacity of glasses
at low temperatures. We present a theory in which we consider the
coupling of the tunnelling motion to nuclear quadrupoles in order
to evaluate the static dielectric susceptibility. In the limit of
weak magnetic field we find the resonant part of the
susceptibility increasing like $B^2$ while for the large magnetic
field it behaves as $1/B$. In the same manner we  consider the
coupling of the tunnelling motion to nuclear quadrupoles and
angular momentum of tunnelling particles in order to find the heat
capacity. Our results show the Schotky peak for the angular
momentum part, and $B^2$ dependence for nuclear quadrupoles part
of heat capacity, respectively.
 We discuss whether or not this
approach can provide a suitable explanation for such magnetic
properties.
\end{abstract}

\pacs{61.43.Fs,77.22.Ch,65.60.+a  }

\maketitle

\section{Introduction}

At very low temperatures, glasses and other amorphous systems show
similar thermal, acoustic and dielectric properties \cite{Esq98},
which are in turn very different from those of crystalline solids.
Below $1 \rm{K}$, the specific heat $C_v$ of dielectric glasses is
much larger and the thermal conductivity $\kappa$ orders of
magnitude lower than the corresponding values found in their
crystalline counterparts. $C_v$ depends approximately linearly and
$\kappa$ almost quadratically on temperature\cite{Zel71}, respectively. This is
in clear contrast to the cubic dependence observed in crystals for
both properties, well understood in terms of Debye's theory of
lattice vibrations. Above $1 \rm{K}$, $C_v$ still deviates
strongly from the expected cubic dependence, exhibiting a hump in
$C_v/T^3$ which is directly related to the so-called boson peak
observed by neutron or Raman vibrational
spectroscopies\cite{Ahm86}. To explain these, it was considered
that atoms, or groups of atoms, are tunnelling between two
equilibrium positions, the two minima of a double well potential.
The model is known as the two level system (TLS) \cite{Phi72,
And72}. In the standard TLS model, these tunnelling excitations
are considered as independent, and some specific assumptions are
made regarding the parameters that characterize them (for a review
see for example Ref.~\onlinecite{Esq98}).

In recent years an intriguing magnetic-field dependence of the
dielectric and coherent properties of some insulating glasses was
reported. In 1998 Strehlow et al. observed a sharp kink at $T=5.84
\; mK$ in the dielectric constant of the multi-component glass
$BaO-Al_2O_3-SiO_2$ \cite{Str98} measured in very weak magnetic
field, of the order of $10 \; \m T$. The effect was several orders
of magnitudes larger than what is expected, considering the absence of
magnetic impurities in this insulator. Later studies carried on in
magnetic field ranging up to $25 \; T$ and temperatures well below
$100 \; mK$ revealed for several materials a complex and strong
dependence of the dielectric response on the external magnetic
field, on the applied voltage and on temperature \cite{Str98,
Str00, Woh01, Lec02, Hau04}. Even more surprising phenomena were
observed in the spontaneous polarization echoes experiments on
$BaO-Al_2O_3-SiO_2$ \cite{Lud02}.
The increase of the echo amplitude by a factor of $4$
was reported when varying the magnetic field
in the range of $0$ to $200 \; mT$. Similar results were later
reported in the case of amorphous mixed crystal $KBr_{1-x}CN_x$
\cite{Ens02}. Until now a unified theory does not exist while some
contradictions are present in the recent works.
We have  clear evidence that such
intriguing magnetic properties are in connection with the low-energy
tunnelling excitations present in almost all amorphous solids.
These tunnelling states are known to be proved very well by
the echo experiments \cite{Wue97}.
On the other hand these
excitations have been reported as showing up at very low temperatures
(usually $T<100 \; mk$), where the TLS are responsible for the
thermal and dynamical properties of glasses \cite{Ens02, Hun00}.

Several generalizations of the standard TLS model have been reported after
the anomalous behavior of glasses in a magnetic field.
The main question for such models is how a TLS should
interact with the magnetic field, being not clear how the
tunnelling entity would acquire a finite magnetic moment.
According to the proposed solutions, the models can be divided into
"orbital" and "spin" models.

In the orbital models, the tunnelling entities are not simple
two-state systems, but performing some kind of circular motions.
Due to the presence of the magnetic field, a charged particle
moves on a loop encloses a magnetic flux and thus can acquire an
Aharonov-Bohm phase. In order to obtain such a closed trajectory,
a "Mexican-hat potential" was proposed by Kettemann, instead of
the usual double-well potential \cite {Ket99}. The resulting flux
$\phi =\pi r^{2}B$, with $r$ being the hat radius, proves to be by
several orders of magnitudes smaller than the flux $\phi
_{0}=h/e$. Even though the existence of cluster configurations of
atoms or molecules containing up to $N=200$ units which contribute
in a collective tunnelling were reported, (see for example
\cite{Lub01} and references therein) it is improbable that
such an effect can be extended to an amount of the order of $N
\sim 10^5$. A detailed discussion of this aspect has been
presented in Ref.\onlinecite{Lan02}. The combination of
the Aharonov-Bohm effect with a flip-flop configuration of the two
interacting TLS has been proposed
to explain the dependence of the echo amplitudes on the applied magnetic field
\cite{Akb05}. A different modality to consider the occurrence of
an apparent flux phase having the correct order of magnitude was
proposed by W\"{u}rger \cite{Wue02}. The mechanism proposed
consist of pair of coupled TLS and a non-linear coupling to the
external voltage. Still, the required degeneracy of the nearby TLS
is not in accordance with the known distribution for the
tunnel splitting. Another possibility for the formation of
closed loops was proposed by Le Cochec \cite{Lec02}, namely a
jagged and uneven potential landscape between the two wells.

To conclude, the "orbital" models can provide an explanation for
some of the magnetic field effects by considering the flux
dependance of the tunnelling splitting. Unfortunately, some
assumptions have been made which cannot be reconciled with the standard
features of the tunnelling model.

The spin models\cite{Wue02a} provide an alternative mechanism for
the observed magnetic properties, especially for the polarization
echoes\cite{Wue04}. A direct coupling between the nuclear spin of
the tunnelling entities and the applied magnetic field is
considered. We can easily notice that the multi-component glasses
used in the echo experiments contain one or several kind of atoms
that carry a nuclear quadrupole. For example in the case of
$BaO-Al_2O_3-SiO_2$ we find the abundant isotopes $^{27}Al(I=5/2)$
and less frequent $^{137}Ba(I=3/2)$; for the boro-silicate we find
the abundant isotope $^{11}B(I=3/2)$. However, the echo
experiments convincingly put into evidence the role of the nuclear
quadrupole moments in glasses. According to this model the
magnetic properties should not be measured in materials whose
nuclei contains no finite quadrupole moment. Until now no
counter-example has been reported. Moreover, recently the role of the
quadrupole moments have been systematically studied and confirmed
by the echo experiments in ordinary glycerol $C_3H_8O_3$ and
deuterated one $C_3D_8O_3$ \cite{Nag04}.

The question, wether or not the nuclear quadrupole should be
responsible for the magnetic field dependence of the dielectric
constant seems to be natural. Relevant experiments involve
temperature where the thermal energy is significantly larger than
the quadrupole splitting. For thermal tunnelling systems, with TLS
splitting ($E$) much larger than the quadrupole splitting, it was
proven already that the resonant, or van Vleck susceptibility is
rather insensitive to the nuclear quadrupole motion \cite{Dor04}.
For such systems  the relaxation or Debye part of the
susceptibility shows a large magnetic field dependence
\cite{Dor06}. By considering TLS with small splitting comparable
to the nuclear quadrupole energies and using a numerical
estimation, a pronounced dependence of the electric permittivity
of the applied magnetic field was obtained \cite{Pol05}.

In this article we have studied the magnetic field dependence of
the dielectric susceptibility and the heat capacity in cold
glasses taking into account the quadrupole effects. The general expressions
in the mentioned cases are complex, however, we have obtained these
thermodynamic behaviour in the small and large magnetic fields limit.
 In Section II we introduce the nuclear
spins in the frame of the TLS. The resulting magnetic field
dependent part of the susceptibility is given in the Section III,
using the perturbation expansion  for two different regimes,
namely small and large magnetic fields. We end this section with a
discussion of our results and a comparison with available
experimental data and previous calculations.
In section IV, the specific heat has been
calculated in  three separate terms, namely TLS part; angular
momentum and nuclear spin parts. Finally the conclusion is
presented with a discussion of our results and a comparison with
available experimental data\cite{Str04}.


\section{TLS with a nuclear spin}

The standard TLS can be described as a particle or a small group
of particles moving in an effective double-well potential (DWP).
At very low temperatures, only the ground states corresponding to
the two wells are relevant. Using a pseudo-spin representation the
Hamiltonian of such a TLS is written
\begin{equation}
H_{TLS} \; = \; \frac{1}{2}\Delta _{0}\sigma _{x} +
\frac{1}{2}\Delta \sigma _{z},
\label{eq1}
\end{equation}
where $\sigma _{z}$ is the reduced two-state coordinate,
\bd \sigma _{z} \; = \; \mid L \ket \bra L \mid - \mid R \ket \bra
R \mid.
\ed
The eigenvalues of $\sigma _{z}$ are $\pm 1$, labelling the localized states in the
two wells (left well $L$, right well $R$), while the tunnelling matrix is
taken into account by
\bd \sigma _{x} \; = \; \mid L \ket \bra R \mid + \mid R \ket \bra
L \mid. \ed
We denoted by $\Delta $ the energy off-set at the bottom of the
wells, and by $\Delta _{0}$\ the tunnel matrix element. According
to the randomness of the glassy structure, the energy difference
between the two wells
\begin{equation}
E \; = \; \sqrt{\Delta _{0}^{2}+ \Delta ^{2}},
\label{eq2}
\end{equation}
have a broad distribution. The energy off-set and the tunnelling
matrix element obey a distribution law
\begin{equation}
{\cal P}\left( \Delta ,\Delta _{0}\right) \; = \;
\frac{P_{0}}{\Delta _{0}},
\label{eq3}
\end{equation}
where $P_0$ is a constant. It is useful to define new spin
operators using the relations
\bea \sigma_x \; = \; u \sigma_E - w \sigma_A, \no \sigma_z \; = \;
w \sigma_E + u \sigma_A,
\label{eq4} \eea
such that the TLS Hamiltonian becomes diagonal
\be H^D_{TLS} \; = \; \frac{E}{2} \; \sigma_E.
\label{eq5} \ee
We used the notations
\be u \; = \; \frac{\Delta_0}{E} \; \; , \; \; w \; = \;
\frac{\Delta}{E}
\label{eq6}
\ee
which satisfy $u^2 + w^2 =1$.

For the moment there is no rigorous theory for tunnelling in
glasses. It is assumed that atoms or groups of atoms participate
in one TLS. As we mentioned before, in the case of the
multi-component glasses, one or several of the tunnelling atoms
carry a nuclear magnetic dipole and an electric quadrupole. When
the system moves from one well to another, the atoms change their
positions by a fraction of an \AA ngstr\"{o}m.

We can describe the internal motion of the nuclei by a nuclear
spin ${\bf I}$ of absolute value ${\bf I}^2=\hbar^2I(I+1)$. This
is related to a magnetic dipole moment $g \mu_N {\bf I}/\hbar$,
where $g$ is the Land\'{e} factor and $\mu _ N \approx 5\times
10^{-27}J/T$ is the nuclear magneton. The magnetic dipole couples
to  an external magnetic field ${\bf B}=B e_B$ and give rise to
a Zeeman term
\be H_Z^{(1)} \; = \; - \epsilon_z e_B\cdot I
\label{eq7}
\ee
where the frequency $\epsilon_z = g \mu_N B /\hbar$ is directly
proportional to $B$.

For a nucleus with spin quantum number $I \geq 1$ the charge
distribution $\rho({\bf r})$ is not isotropic. Besides the charge
monopole, an electric quadrupole moment can be defined with
respect to an axis ${\bf e}$
\begin{equation}
Q=\int d^{3}r\left[ 3({\bf r\cdot e})^{2}-{\bf r}^{2}\right] \rho
({\bf r}).
\label{eq8}
\end{equation}
This can couple to an electric field gradient (EFG) at the nuclear
position, expressed by the curvature of the crystal field
potential. The potential describing this coupling reads as
\cite{Abr89}
\begin{equation}
V_Q \; = \;
\frac{-eQ}{I(2I-1)}[V_{11}I_{1}^{2}+V_{22}I_{2}^{2}+V_{33}I_{3}^{2}].
\label{eq9}
\end{equation}
The bases used here $(e_1,e_2,e_3)$ are the principal axes of the
tensor $V_{ij}$ which describes the electric field gradient, and
$e$ is the electron charge. According to the Laplace equation the
potential obey $V_{11}+V_{22}+V_{33}=0$. If we define the
asymmetry parameter $\eta = \frac{V_{22}-V_{33}}{V_{11}}$, the
quadrupole potential can be expressed as:
\begin{equation}
V_Q \; = \;\epsilon_q[3I_{1}^{2}+\eta (I^{2}_{2}-I_{3}^{2})-I^{2}]
\label{eq10}
\end{equation}
where we denote by $\epsilon_q=\frac{-eQ V_{11}}{4I(2I-1)}$ the
quadrupole coupling constant.


\section{Static dielectric susceptibility}

Our purpose is to determine the magnetic field dependent part of
the static dielectric susceptibility for a TLS coupled with a
nuclear quadrupole.
Therefore we have to consider the interaction of the dipole
operator $(1/2)P \sigma_z$ with an external electric field $F$.
The dipole moment arises from the relative motion of partial
charges related to the atoms forming the tunnel system. This
interaction will be described by
\be  V_F=(P \cdot F)\sigma_z.
\label{eq11}
\ee
 If we denote
\begin{equation}
Z \; = \; {\rm Tr}\left\{ e^{-\beta H}\right\},
\label{eq12}
\end{equation}
the partition function of the system, where $H$ is the total
Hamiltonian; the trace, denoted later by $<...>$ involves
two-state and spin variables. Here
 $\beta =
\frac{1}{ K_BT}$, $K_B$ is the Boltzmann constant and $T$ is
temperature. We can express the susceptibility as
\begin{equation}
\widehat{\chi}\; = \; \overline{-\left\langle \frac{\partial
^{2}f}{\partial F^{2}}\right\rangle},
\label{eq13}
\end{equation}
where $f=-\frac{1}{\beta }\ln Z$ represents the free energy. The
statistical average implies also an integration over the two
parameters of the TLS, the energy off-set and the tunnelling
matrix element, according to the distribution law given by Eq.
(\ref{eq3}), as well as an integration over the nuclear quadrupole
parameters; this step, denoted by $\overline{(...)}$, will be
discussed at the end of the calculation. We can write the partition
function in terms of the energy levels $E_{i,m}$ (eigenvalues of
the total Hamiltonian)
\be Z=\sum_{i=L/R} \sum_{m=-I}^Ie^{-\beta E_{i,m}}.
\label{eq14}
 \ee
Therefore, the susceptibility contains three terms:
\bea \chi^{(1)} &=& -\frac{1}{Z}\sum_{i,m}e^{-\beta
E_{i,m}}\frac{\partial ^{2}}{\partial F^{2}} E_{i,m}, \no
\chi^{(2)} &=& \frac{\beta}{Z}\sum_{i,m}e^{-\beta E_{i,m}}
\left[\frac{\partial }{\partial F} E_{i,m}\right]^{2}, \no
\chi^{(3)} &=&-\frac{\beta}{Z^2}\left[\sum_{i,m}e^{-\beta E_{i,m}}
\frac{\partial }{\partial F} E_{i,m}\right]^{2}.
\label{eq15} \eea
%
The eigen energies ($E_{i,m}$) of the full Hamiltonian can not be obtained analytically
in the general case because the Zeeman and nuclear quadrupole terms do not commute.
In this respect we will consider two different regimes where we can obtain the
analytic results in a perturbation frame work.
In order to develop a perturbation expansion we will consider
two separate cases, a small external magnetic field
and a large one, respectively.

\subsection{Small magnetic field}
We begin with the case in which the Zeeman term is smaller than
the nuclear quadrupole potential, $\varepsilon_z\ll
\varepsilon_q$, thus the Zeeman term is  treated as a weak
perturbation. If we consider the simplest symmetric case,
$\eta=0$, then the nuclear quadrupole potential in the $I_1$ basis
has the following representation
\bea H_q=\left[ \varepsilon_q^L{\cal
V}_{I,m}(\frac{1+\sigma_z}{2})+ \varepsilon_q^R{\cal
V}_{I,m}(\frac{1-\sigma_z}{2})\right]|I,m\ket \bra I,m|, \no
\label{eq16}
\eea
where we have defined  ${\cal V}_{I,m}=3m^2-I(I+1)$, and
$\varepsilon_q^{L/R}$ as the quadrupole coupling constant in the
left and the right well.
Therefore, in the $\sigma_E$ basis we can write
\bea H_q=\left[ \varepsilon_q{\cal V}_{I,m} + \varepsilon_q'{\cal
V}_{I,m}(w \sigma_E+u\sigma_A)\right]|I,m\ket \bra I,m|.
\label{eq17}
\eea
Here we have defined $\varepsilon_q=\frac{1}{2}({ \varepsilon_q^L+
\varepsilon_q^R})$ and $\varepsilon_q'=\frac{1}{2}({
\varepsilon_q^L- \varepsilon_q^R})$.
We can split the total Hamiltonian in two parts,
\bea H=H^{(0)}+\delta H,
\label{eq18}
\eea
where the unperturbed part is
\bea H^{(0)}=\left[\frac{E}{2}\sigma_E+\varepsilon_q{\cal
V}_{I,m}-m\varepsilon_z\cos\theta'\right]|I,m\ket \bra I,m|,
\label{eq19}
\eea
and the perturbation term is
\bea \delta H&=&\left[( \varepsilon_q'{\cal V}_{I,m}+PF)(w
\sigma_E+u\sigma_A)\right]|I,m\ket \bra I,m|
\no &-& \alpha_{-}^{m'}\varepsilon_z |I,m\ket \bra I,m+1|
\no&-&\alpha_{+}^{m'}\varepsilon_z |I,m\ket \bra I,m-1|.
\label{eq20}
\eea
We have defined
\bea \alpha_{\pm}^{m'}=\sqrt{I(I+1)-m(m\pm 1)}\sin\theta'e^{-[\pm
i \phi' ]},
\label{eq21}
 \eea
where $\theta'$ and $\phi'$ are the direction of magnetic field
in the basis  $(e_1, e_2, e_3)$; here we have assumed that $e_1=e_z$.

We apply the perturbation theory to get the second order correction for the
energy levels of the system, $E_{i,m}$. Therefore, the first
order correction is
\bea \delta E^{(1)}_{i,m}\; = \; w(P\cdot F+\varepsilon_q'{\cal
V}_{I,m})\bra i |\sigma_E|i\ket.
\label{eq22}
\eea
For the sake of simplicity we use $\bra i |\sigma_E|i\ket$
instead of $\bra i,m |\sigma_E|i,m\ket$, which is independent of the
magnitude of $m$. The second order correction of the energy reads
\bea &&\delta E_{i,m}^{(2)}= \sum_{j,n|(j,n)\neq(i,m)} \frac{|\bra
i,m |\delta H|j,n\ket|^2}{E^{(0)}_{i,m}-E^{(0)}_{j,n}}=
 \no&&\epsilon_z^2
\left[
 \frac{|\alpha^{'m}_{+}|^2}{(2m-1)\varepsilon_q-\varepsilon_z\cos\theta'}\right.-
\left.\frac{|\alpha^{'m}_{-}|^2}{(2m+1)\varepsilon_q-\varepsilon_z\cos\theta'}\right]
\no&&+ \frac{u^2}{E}(P\cdot F+\varepsilon_q'{\cal V}_{I,m})^2\bra
i |\sigma_E|i\ket,
\label{eq23}
\eea
where $E^{(0)}_{i,m}$ is the eigenvalue of the unperturbed part of
the Hamiltonian $H^{(0)}$. Taking into account the second order
correction of energy, the susceptibility is calculated by
Eq.(\ref{eq15}). By using the following relations
\bd \frac{\partial }{\partial F} E_{i,m}\vert_{F=0} \; = \; wP\bra
i |\sigma_E|i\ket,
\label{eq24}
\ed
and
\bea \frac{\partial ^2}{\partial F^2} E_{i,m}\vert_{F=0}=
\frac{2P^2u^2}{E}\bra i |\sigma_E|i\ket,
\label{eq24b}
\eea
we can immediately express the second term of the susceptibility
\be \chi_{2}\;=\;\beta P^2w^2.
\label{eq25}
\ee
The third term is
\be \chi_{3}\;=\; -P^2w^2\beta t^2,
\label{eq26}
\ee
where we defined  $t=\tanh(\beta E/2)$. The first
term of the susceptibility contain the second derivative of the
energy levels with respect to the electric field. After some simple
algebra $\chi_1$, can be written as
\bea \chi_{1}&=& \frac{2P^2u^2}{ZE} \sum_{m} \left[e^{-\beta
E_{d,m}}-e^{-\beta E_{u,m}}\right],
\label{eq27}
\eea
where $Z=\sum_{m} \left[e^{-\beta E_{d,m}}+e^{-\beta
E_{u,m}}\right]$. Defining $\Lambda_m$ as
\bea &&\Lambda_m=\varepsilon_q{\cal
V}_{I,m}-\varepsilon_zm\cos\theta' + \no
&&
 \epsilon_z^2
\left(
 \frac{|\alpha^{'m}_{+}|^2}{(2m-1)\varepsilon_q-\varepsilon_z\cos\theta'}
 -\frac{|\alpha^{'m}_{-}|^2}{(2m+1)\varepsilon_q-\varepsilon_z\cos\theta'}\right),
\nonumber
\label{eq27b}
\eea
one can write that
\bea E_{i,m}\vert_{F=0}=\left[\frac{E}{2}+
 \frac{u^2}{E}(\varepsilon_q'{\cal
V}_{I,m})^2\right]\bra i |\sigma_E|i\ket+\Lambda_m. \;\;\;\;\;\;%
\label{eq28}
\eea
The partition function can be written as $ Z=\sum_{m}
 \cosh
\left[\frac{\beta E}{2}+
 \frac{u^2\beta}{E}(\varepsilon_q'{\cal
V}_{I,m})^2\right]
 e^{-\beta \Lambda_m}$.
Then the first term of the susceptibility becomes
\bea \chi_{1}&=& \frac{2P^2u^2}{ZE} \sum_{m}
 \sinh
\left[\frac{\beta E}{2}+
 \frac{u^2\beta}{E}(\varepsilon_q'{\cal
V}_{I,m})^2\right]
 e^{-\beta \Lambda_m}.
 \no
\label{eq29}
 \eea
If at this point we assume $\varepsilon'_q=0$ and $B=0$ we will simply
recover the result of standard TLS model, $\chi_{1}=
\frac{2P^2u^2t}{E}$.

\subsubsection{Thermal expansion}

The experiments we are addressing are performed at temperatures
between few tens and few hundreds of milliKelvin, where the
thermal energy $k_BT \sim 10^{-24} J$ is few orders of magnitude
larger than the Zeeman term (at $B=1T$ for example $\mu_N B/k_B
\sim 0.4 mK$) or the quadrupole coupling (echo experiments
suggested that $\epsilon_q \sim \epsilon_z$ at $200mT$, so smaller
than one $mK$). We can conclude that the parameters $\beta
\epsilon_z$, $\beta \epsilon_q$  are small, which justify an
expansion of $\chi_{1}$ and the exponential factors like $e^{\pm
\beta \epsilon_{z,q}}$. Carrying out such an expansion in the
thermal energy,
 we can
easily show that
%
%
\bea
&&\Delta\chi_{1}(B)\simeq\frac{2P^2u^4}{(2I+1)E^2}
(1-t^2)\varepsilon_q'^2
 \beta^2 \epsilon_z^2\times
 \no
 &&\sum_{m}
\left[
 \frac{|\alpha^{'m}_{+}|^2}{(2m-1)\varepsilon_q-\varepsilon_z\cos\theta'}
 -\frac{|\alpha^{'m}_{-}|^2}{(2m+1)\varepsilon_q-\varepsilon_z\cos\theta'}\right].
 \no
\label{eq30}
 \eea
%

In order to complete the calculation of the magnetic field
dependent part of the static susceptibility we need to perform the
integral over the TLS parameters, considering the distribution
function defined in Eq. (\ref{eq3}), for this regime
($\varepsilon_z\ll \varepsilon_q$), which can be rewritten as
\bea \Delta\chi_{1}(B)\simeq \frac{2P^2\kappa\xi}{(2I+1)} \beta^3
\varepsilon_q'^2 \frac{\epsilon_z^2}{\varepsilon_q}
\label{eq31}
\eea
 where $\kappa=\left[\frac{1}{2I+1}
(\sum_{m}{\cal V}_{I,m}^2)(\sum_{m}\alpha_m )- \sum_{m}{\cal
V}_{I,m}^2\alpha_m\right]$, $\alpha_m=\sum_{m}[
\frac{|\alpha^{'m}_{+}|^2}{(2m-1)}
 -\frac{|\alpha^{'m}_{-}|^2}{(2m+1)}]$, and $\xi$
 is a constant (nonmagnetic-dependent) which
  comes from averaging over the TLS parameters which depends on
 the magnitude of  the lower
and upper bound of the two-level splitting, $\Delta_{0min}$ and
$E_{max}$. It can be easily shown that in the case of same EFG in
the two wells ($\e_q^L = \e_q^R$), the dependence of the static
dielectric susceptibility on the magnetic field will vanish.


\subsection{Large magnetic field}

 Let us denote by $(e_1, e_2, e_3)$ the basis
in the left well and by $(e_1^{\prime}, e_2^{\prime},e_3^{\prime})$ the
corresponding one in the right well. If we suppose that
$e_3^{\prime}\parallel e_3\parallel \hat{z}$ and $e_1^{\prime},e_1$
make an angle $\theta$ with each other \cite{Pol05}, we can
write the total Hamiltonian in the basis of the operator $I_3=I_z$
in the following form
\bea
 \left(\begin{array}{cc} H^L+ \left( \frac{\Delta}{2}+PF
\right)\cdot\textbf{1
}&\frac{\Delta_0}{2}\cdot\textbf{1 }\\
\frac{\Delta_0}{2}\cdot\textbf{1 }&H^R-\left( \frac{\Delta}{2}+PF
\right)\cdot\textbf{1 }\end{array}\right) \vspace{0.5cm}
\label{eq32}
\eea
where $\textbf{1}$ is the unit matrix of rank $(2I+1)$.
$H^L=V_Q^L+H_Z^{(1)L}$ and $H^R=V_Q^R+H_Z^{(1)R}$ are defined for
particles in the left and right wells, respectively. The state in
each well (L or R) can be characterized by $\mid
\psi^{L(R)}_{I,m}\ket=\mid L(R)\ket\bigotimes \mid I,m\ket$ and
consequently the matrix element of the Hamiltonian is defined by
the following equation,
\begin{equation} \label{aaaa} H^{L(R)}_{m,n}=\bra
\psi^{L(R)}_{I,m}\mid H^{L(R)}\mid \psi^{L(R)}_{I,n}\ket.
\label{eq33}
\end{equation}
After some simple algebra (see for example
Ref.~\onlinecite{Akb06}) we can easily show that for $H^{R}_{m,n}$
we obtain
\begin{eqnarray} H^{R}_{m,n}&=&
-\epsilon_z\cos\theta^{''}
m\delta_{m,n}+\epsilon_q^R\delta_{m,n}\Upsilon_{m,m} \no &
-&\epsilon_z\sin\theta^{''}
\left[e^{i\phi^{''}}\alpha_{-}^m\delta_{m+1,n}+e^{-i\phi^{''}}\alpha_{+}^m\delta_{m-1,n}\right]
  \no
&+&\frac{\epsilon_q^R}{4}\delta_{m-2,n}
\Upsilon_{m,m-2}e^{2i\theta}
+\frac{\epsilon_q^R}{4}\delta_{m+2,n}\Upsilon_{m,m+2}e^{-2i\theta},
\no
\label{eq34}
\end{eqnarray}
where $\theta^{''}$ and $\phi^{''}$ define the direction of
magnetic field in the basis  $(e_1, e_2, e_3)$; $\alpha_{-}^m$ and
$\alpha_{+}^m$ are denoting
\bea \alpha_{\pm}^m=\sqrt{I(I+1)-m(m\pm 1)},
\label{eq35}
 \eea
and
\bea \Upsilon_{m,m}&=&\left[ \frac{1}{2} I(I+1)-\frac{3}{2}m^2
+\frac{\eta }{2} [I(I+1) -3m^2]\right],
\no \Upsilon_{m,m-2}&=&\frac{1}{4}\left[(3-\eta)
[I(I+1)-(m-1)(m-2)]^{\frac{1}{2}}\right.
\no&\times&\left.[I(I+1)-m(m-1)]^{\frac{1}{2}}\right], \no
\Upsilon_{m,m+2}&=& \frac{1}{4}\left[(3-\eta)
[I(I+1)-(m+1)(m+2)]^{\frac{1}{2}}\right. \no
&\times&\left.[I(I+1)-m(m+1)]^{\frac{1}{2}}\right].
\label{eq36}
\eea

In a similar manner we can obtain the matrix element of the
Hamiltonian for the particles in the left well $H^{L}_{m,n}$ by
taking $\theta=0$ and changing
$\epsilon_q^R\longmapsto\epsilon_q^L$ in Eq.~(\ref{eq34}).

We will use the perturbation approach assuming that $\varepsilon_z
\gg \varepsilon_q$ (here for satisfying the perturbation condition
we consider the special magnetic direction in a way that
$\theta^{''}\ll \frac{\pi}{2}$). It proves to be useful to change
the representation of the Hamiltonian into the basis of the spin
operators $\sigma_E$ and $\sigma_A$. In this basis we split the
total Hamiltonian of system as $H_0^D+\delta V$, where $H_0^D$,
the unperturbed part, will now contain the diagonal matrix
elements of the total Hamiltonian except those coming from $V_F$;
therefore the perturbation, $\delta V $, will contain the
non-diagonal terms of the Hamiltonian together with $V_F$. We find
that
\bd H_0^D \; = \; \sum_{m=-I}^I \left[\frac{E}{2}\; \sigma_E
 -(m\epsilon_z\cos\theta{''}) \textbf{1} \right] |I,m\ket \bra I,m|,
\label{eq36b}
\ed
and
\bea \bra m |\delta V|n\ket \; = \; \varepsilon _{mn}+\eta _{mn}w
\sigma _{E}+\eta _{mn}u\sigma _{A}.
\label{eq37}
\eea
We denote $\varepsilon _{mn}^{L/R}=\langle m|\delta
V^{L/R}|n\rangle $ and
\bea \varepsilon _{mn} = \frac{1}{2}\left( \varepsilon
_{mn}^{L}+\varepsilon _{mn}^{R}\right) ,\;\;\; \eta _{mn}=
\frac{1}{2}\left( \varepsilon _{mn}^{L}-\varepsilon
_{mn}^{R}\right).
\label{eq38}
 \eea
The diagonal part of the perturbation term is
\be V^D_F \; = \; (w\sigma_E+u\sigma_A) P \cdot F.
\label{eq39}
 \ee
The first order correction of the energy level will be obtained as
\be  \delta E_{i,m}^{(1)}\;=\; w  P \cdot F\bra i |\sigma_E|i\ket+
\varepsilon_{mm}+\eta_{mm} w \bra i |\sigma_E|i\ket,
\label{eq40} \ee
and the second order one will be
\bea
 &&\delta E_{i,m}^{(2)}=
 P^2 F^2 u^2\sum_{j\neq i}\frac{1}{E^{(0)}_{i,m}-E^{(0)}_{j,m}} \no &+&
PFu^2\sum_{j\neq i}\left[\frac{|\eta_{mm}\bra j |
\sigma_A|i\ket|^2}{E^{(0)}_{i,m}-E^{(0)}_{j,m}} +C.C. \right]
 \no &+&
 \sum_{j\neq i}
 \frac{|\eta_{mm}|^2u^2}{E^{(0)}_{i,m}-E^{(0)}_{j,m}}
 +\sum_{n\neq m}
 \frac{|\bra i |
\sigma_E|i\ket\eta_{mn}w+\varepsilon_{mn}\textbf{1}|^2}{E^{(0)}_{i,m}-E^{(0)}_{i,n}}.
\no
\label{eq41}
\eea
Since we have assumed that $\delta V$ do not contain any diagonal
term, $\varepsilon_{mm}=\eta_{mm}=0$, the corrections to the
energy levels simplify to
\be  \delta E_{i,m}^{(1)}\;=\; w  P \cdot F\bra i |\sigma_E|i\ket,
\label{eq42} \ee
and
\bea
 \delta E_{i,m}^{(2)}&=& \frac{ P^2 F^2 u^2}{E}\bra i |\sigma_E|i\ket
 \no &+  & \sum_{n\neq m}
 \frac{|\bra i |
\sigma_E|i\ket\eta_{mn}w+\varepsilon_{mn}\textbf{1}|^2}{E^{(0)}_{i,m}-E^{(0)}_{i,n}}
.
\label{eq43}
\eea

For the large  magnetic field regime we calculate the first and
the second derivatives of $E_{i,m}$ with respect to the electric
field
\bea \frac{\partial }{\partial F} E_{i,m}\vert_{F=0}= w  P \bra i
|\sigma_E|i\ket,\nonumber
\label{eq44a}
\eea
and
\bea \frac{\partial ^2}{\partial F^2} E_{i,m}\vert_{F=0}=
\frac{2P^2 u^2}{E}\bra i |\sigma_E|i\ket.
\label{eq44b}
\eea
We can easily express the susceptibility
\bea \chi^{(1)} &=&\frac{2P^2 u^2 }{ZE}\sum_{m}[e^{-\beta
E_{d,m}}-e^{-\beta E_{u,m}}].
\label{eq45}
 \eea
Here  $Z = \sum_{m}[e^{-\beta E_{d,m}}+e^{-\beta E_{u,m}}]$ is the
partition function, and $E_{d,m}$ and $ E_{u,m}$ are defined by
\bea E_{i,m}|_{F=0}& =& \frac{E}{2}\bra i|\sigma_E|i\ket
-m\cos\theta^{''} \varepsilon_{z} \no&+& \sum_{n\neq m}
 \frac{|\bra i |
\sigma_E|i\ket\eta_{mn}w+\varepsilon_{mn}\textbf{1}|^2}{E^{(0)}_{i,m}-E^{(0)}_{i,n}}.
\label{eq46}
 \eea
%

\subsubsection{Thermal expansion}
Let us remember here that the thermal energy exceeds  few
orders of magnitude the Zeeman energy and the nuclear quadrupole
potential. For this reason we can carry out a thermal expansion in
terms of $\beta \varepsilon_q$, $\beta \varepsilon_z$ and find
\bea &&e^{-\beta E_{i,m}}=e^{-\beta ( \frac{E}{2}\bra
i|\sigma_E|i\ket)}\times \no&&[1+m\cos\theta^{''}
\beta\varepsilon_{z} - \beta\sum_{n\neq m}
 \frac{|\bra i |
\sigma_E|i\ket\eta_{mn}w+\varepsilon_{mn}\textbf{1}|^2}{-\cos\theta^{''}
\varepsilon_{z}(m-n)}]\no&&+O(\beta^2\varepsilon^2),
\label{eq47}
 \eea
After some algebra, we can easily write
%
\bea
 \chi^{(1)}&=&\frac{2P^2 u^2}{E}t +
 \frac{2\beta P^2
u^2}{E(2I+1)}(1-t^2)\times
 \no&&
\sum_{m,n\neq m}
 w\frac{\eta_{mn}^{*}\varepsilon_{mn}+\eta_{mn}\varepsilon_{mn}^{*}}{\cos\theta^{''}
\varepsilon_{z}(m-n)}+ O(\beta^2\varepsilon^2).
\label{eq48}
\eea
%

Similar to the small magnetic field regime, from Eq.~(\ref{eq15})
and Eq.~(\ref{eq44b}) we can straightforwardly show that
\bea
 \chi^{(2)}+ \chi^{(3)}&=&P^2w^2\beta(1-t^2),
\label{eq49}
\eea
The magnetic field dependent correction of the susceptibility will
be
\bea
 \delta\chi(B)&=&
 \frac{4\beta P^2
}{E(2I+1)}(1-t^2)wu^2 \sum_{m,n\neq m}
 \frac{\rm{Re}[\eta_{mn}^{*}\varepsilon_{mn}]}{\cos\theta^{''}
\varepsilon_{z}(m-n)}
\no&+& O(\beta^2\varepsilon^2).
\label{eq50}
 \eea
If we set $ \varepsilon_q=0$, we can assume that $I=0$, we
recover once more the result of standard TLS model,
\bea
 \chi_{TLS}&=&\frac{2P^2 u^2}{E}t+P^2w^2\beta(1-t^2).
\label{eq51}
\eea
%

By averaging over the TLS parameters according to the distribution
function (\ref{eq3}), we obtain:
\bea
 \delta\chi(B)
&=&
 \frac{4\beta P^2\xi'
}{(2I+1)} \sum_{m,n\neq m}
 \frac{\rm{Re}[\eta_{mn}^{*}\varepsilon_{mn}]}{\cos\theta^{''}
\varepsilon_{z}(m-n)}
\no&+& O(\beta^2\varepsilon^2).
\label{eq52}
 \eea
where $\xi'$ is the numeric constant  (nonmagnetic-dependent)
which depends on the magnitude of $E_{max}$ and $\Delta_{0min}$.
We can easily see that $\delta\chi(B) \sim \frac{\beta
\bar{\varepsilon_q^2}}{\varepsilon_z}$, where
$\bar{\varepsilon_q^2}\propto
[(\varepsilon^L_q)^2-(\varepsilon^R_q)^2]$.

As we expected   from the beginning  if we neglect the phase
difference between the nuclear moment in the wells and if we
assume that the EFG in the both wells are the same
$(\varepsilon_q^L=\varepsilon_q^R)$, then we do not find any
magnetic dependence in the second order perturbation
(Eq.~\ref{eq31} and Eq.~\ref{eq52}).


\subsection{Discussion on the field dependence of dielectric susceptibility }

In this section we have addressed the question whether or not the
coupling of the two-state coordinate and the nuclear spin
variables through the nuclear quadrupole potential in an
inhomogeneous crystal field can been taken into account as the
source of the magnetic field dependence of the dielectric
susceptibility.

Analyzing the existent data regarding the dielectric
susceptibility as a function of the magnetic field of different
oxide glasses and mixed crystals
\cite{Str98,Str00,Woh01,Lec02,Hau04}, we can notice along with a
pronounced bump around $200 mT$, some irregular oscillations. The
curvature of the susceptibility changes its sign few times up to
$5 T$ of the magnetic field. The amplitude of the real part of the
susceptibility seems to be about 10 percent of the TLS
susceptibility, corresponding to approximately $10^{-4}$ of the
dielectric constant. Relevant experiments involve temperatures
from tens of milliKelvin to a few hundreds. We can observe that in
this range, the dielectric constant vary with the inverse
temperature, $1/T$.

The present work reconsider the static dielectric susceptibility of a glassy
systems in an external magnetic field implementing a perturbation approach for the
energy levels . Starting from the coupling of
the nuclear quadrupole with the tunnelling system, calculations
have been performed for two different regimes, i.e. small and
large external magnetic fields with respect to the nuclear
quadrupole potential. We have found that the magnetic field dependent part
of the susceptibility in both regimes is as the following
\bea \Delta \chi_{small}(B) &\sim& \beta^2 \epsilon_z^2
\frac{\epsilon_q'^2}{\epsilon_q}, \no
\Delta \chi_{large}(B)
&\sim& \beta \frac{\bar{\epsilon_q^2}}{\epsilon_z}. \nonumber
\label{eq53a}
 \eea
As it is obvious from the above results the dielectric
susceptibility depends on the magnetic field in the second order
correction of the perturbation scheme via the Zeeman and the
nuclear quadrupole terms. This correction is directly the result
of different electric field gradients (EFG) in the two wells. In
the small magnetic field regime the correction increases with
$B^2$ while it decreases inversely with field ($1/B$) in the large
magnetic field. The dependence on the temperature was put into
evidence in agreement with the existent experimental data. The
magnetic field dependence will disappear in the second order
corrections if we take the same EFG in both wells. We would like
to remind that the previous existing perturbation calculations
\cite{Dor04} which has been considered the same EFG provides a
magnetic field dependence of the dielectric susceptibility only on
the fourth order of expansion.

The theoretical approach we have used do not allow us to calculate
the static dielectric susceptibility for the case where the
applied magnetic field has such values so that the Zeeman energy
and the nuclear quadrupole have similar magnitudes, $\epsilon_z
\simeq \epsilon_q$. But for this regime we know that \cite{Dor04}
the magnetic field dependent part of the static susceptibility
$\chi$ is proportional to the small ratio $[(\varepsilon_z
\varepsilon_q)/T^3]^2$,
 this ratio being of the order of
$10^{-6}$ (one million part) for $T \sim 1 K$.

We have to mention here that our calculation provides only the
static limit ($\omega = 0$) of this contribution.

The dynamical susceptibility of a simple TLS was expressed by
J\"{a}ckle \cite{Jac72}
\bea \chi_{TLS}^{rel}(\omega) \; = \; \left< P^2 w^2
\frac{\beta}{\cosh^2(\beta E/2)} \frac{i \gamma}{\omega + i
\gamma} \right>,
\label{eq53}
\eea
where $\gamma$ is the relaxation rate. Now, by considering the
coupling of the nuclear quadrupole to the tunnelling motion, the
dynamic susceptibility proves to be more complicated, but it can be
generally expressed as \cite{Dor06}
\bea \chi^{rel}(\omega) \; = \; \left< P^2 w^2
\frac{\beta}{\cosh^2(\beta E/2)} \sum_m A_m \frac{i r_m}{\omega +
i r_m} \right>
\label{eq54} \eea
where $r_m$ represents  the eigenvalues of the relaxation rate
matrix, and $A_m$ the corresponding amplitudes ($m=0, \ldots ,2I+1
$). In the limit of $\omega \rightarrow 0$, an approximation of
the susceptibility is provided by Eqs. (\ref{eq31}) and
(\ref{eq52}) for the two considered regimes. Reminding that the
relevant experiments are performed on frequencies in the range of
kilohertz, and that the relaxation rates should be extremely small
quantities (details will be give elsewhere \cite{Dor06}). We can
conclude that such a limiting case cannot provide a proper
explanation for the existent experimental data.

The method we developed cannot provide a plausible amount for the
relative magnitude of the magnetic field dependence part with
respect to the simple TLS susceptibility. We do not find
oscillations of the dielectric susceptibility at magnetic fields
of the order of $1T$.

\section{The heat capacity}
Our purpose is to find the magnetic field dependence of the
heat capacity in multicomponent glasses. To the best of our knowledge there
are not many experimental data for heat capacity of (nonmagnetic)
glasses in the presence of magnetic field. Therefore, we
assume that the tunneling particles have an angular momentum, $J$,
(intrinsic angular momentum such as electron spin). Then, we
must add an additional term to the Hamiltonian which in the presence
of magnetic field leads to
\be H_Z^{(2)} \; = \; - \epsilon_z' J_z.
\label{eq55}
 \ee
\noindent The energy scale is  $\epsilon_z' = g_E \mu_B B /\hbar$,
$\mu_B$ is the Bohr magneton and $g_E$ is the electronic
Land\'{e} factor. Here we have neglected the spin-spin interaction
in the Hamiltonian. We have also assumed that the quantization
axes of angular momentum, $J_z$,  is in the direction of $e_B$
(for more detail, see the Appendix).
The total Hamiltonian of the system is then
\bd H \; = \; H_{TLS}\; + \; H_{Z} + \;H_Q . \ed
Here $H_Z= H_Z^{(1)}+H_Z^{(2)}$ (defined in Eqs.~(\ref{eq7},
\ref{eq55})) and \be
H_Q=(\frac{1+\sigma_z}{2})V_Q^L+(\frac{1-\sigma_z}{2})V_Q^R,
\label{eq56}
 \ee
where $V_Q^{R(L)}$ is defined in Eq.~(\ref{eq10}) for the
particles in right (left) well \cite{Dor04}.
%
%
%
The heat capacity is expressed by the following
relation
\begin{equation}
C_v =\frac{1}{K_BT^2}\left\langle \frac{\partial ^{2}\ln
Z}{\partial \beta^{2}}\right\rangle,
\label{eq58}
\end{equation}
where $Z$ is the partition function which is a sum over the TLS, spin and
angular momentum degrees of freedom.
If we  neglect the phase difference between the nuclear moments in
the two wells (taking $\theta=0$) and  assume that the EFG in
both wells are the same ($\epsilon_q^R=\epsilon_q^L=\epsilon_q$),
then the partition function can be written in the following form
\begin{equation}
Z \; = \; Z_{TLS}SS',
\label{eq59}
\end{equation}
where
\bea Z_{TLS}&=& \; {\rm Tr}\left\{ e^{-\beta
H_{TLS}}\right\}=2\cosh(\beta E/2),
 \no S&=& \; {\rm Tr}\left\{
e^{-\beta (H_z^{(1)}+H_Q)}\right\},
 \no  S'&=& \; {\rm
Tr}\left\{ e^{-\beta H_z^{(2)}}\right\}.
\label{eq60}
 \eea

Therefore, according to Eq.~(\ref{eq58}) and Eq.~(\ref{eq59}),  we
 find that
 \bea C_v& =&\frac{1}{K_BT^2}\left\langle
\frac{\partial ^{2}\ln Z_{TLS}}{\partial \beta^{2}}+\frac{\partial
^{2}\ln S}{\partial \beta^{2}}+\frac{\partial ^{2}\ln S'}{\partial
\beta^{2}}\right\rangle \no &=&C_{TLS}+C^{(I)}+C^{(II)}.
\label{eq61}
\eea By some calculations and after averaging over TLS parameters,
we can easily show that [see for example
Ref.~(\onlinecite{Esq98})]
\bea C_{TLS}&=&\frac{\pi^{2}}{6}P_0K_{B}^{2}T
 \propto T,
\label{eq62}
 \eea
which is independent of the magnetic field. The Zeeman
contribution of the tunneling particle's angular momentum ($J$) in
the partition function is simplified to the following form,
\bea S'=\sum_{m'=-J}^Je^{\beta m'\epsilon'_z}= \frac{\sinh[\beta
\epsilon'_z (J+\frac{1}{2})] }{\sinh[ \frac{\beta
\epsilon'_z}{2}]}.
\label{eq63}
\eea
Thus the Zeeman contribution of the angular momentum in the
specific heat is given by
\bea &&C^{(II)}=\frac{P'}{K_BT^2}\left\langle \frac{\partial
^{2}\ln S'}{\partial \beta^{2}}\right\rangle= \frac{K_B \beta^2
\varepsilon_z'^2}{4}\times P'
 \no&&\left[ csch^2\left(\frac{\beta
\varepsilon_z'}{2}\right)-(2J+1)^2csch^2\left(\frac{\beta
\varepsilon_z'(2J+1)}{2}\right)\right]. \no
\label{eq64}
\eea
where $P'$ is the concentration of intrinsic angular momentum times a
constant which comes from the averaging over the TLS parameters.

Finally, the contribution of the remaining term of the Heat Capacity
to the specific heat, $C^{(I)}$,  is given by

\bea C^{(I)}=K_B\beta^2\left\langle\frac{\partial ^{2}\ln
S}{\partial \beta^{2}}\right\rangle,
\label{eq65}
\eea
where $S=Tr\{e^{-\beta H'}\}$ and $H'=H_z^{(1)}+H_Q$.

As we have mentioned in previous section the thermal energy exceeds
few orders of magnitude the Zeeman energy and the nuclear
quadrupole potential. Therefore, $\beta \epsilon_z$ and $\beta
\epsilon_q$ are very small quantities, consequently $\beta  H'
\ll 1$. Therefore, Eq. (\ref{eq65}) is approximated by
\bea C^{(I)}&=&K_B\beta^2\frac{\partial ^{2} }{\partial
\beta^{2}}\ln\left[ Tr( 1-\beta H'+\frac{\beta^2}{2}H'^2+\ldots
)\right]
\no
&=&K_B\frac{\beta^2}{2I+1}\left[Tr(H'^2)-\frac{Tr(H')^2}{2I+1})
\right]+O(\beta^3\varepsilon^3),\no
\label{eq66}
\eea
and finally we arrive at the following expression
\bea
C^{(I)}&=&P''K_B\beta^2\left[\gamma_1\varepsilon_z^2+\gamma_2\varepsilon_q^2\right]+O(\beta^3\varepsilon^3).
\label{eq67}
 \eea
where $P''$ is the concentration of the nuclear spin times a constant
which comes from the averaging over the TLS parameters. Here
$\gamma_1$ and $\gamma_2$ are numerical constant which have been
defined by~(see appendix A),
\bea \gamma_1&=&\frac{1}{2I+1}\sum_m\left[m^2\cos^2\theta^{'}
+\sin^2\theta^{'}[I(I+1)-m^2] \right],
\no
\gamma_2&=&\frac{1}{2I+1}\sum_m[\Upsilon_{m,m+2}^2+\Upsilon_{m,m-2}^2-\frac{\Upsilon_{m,m}^2}{2I+1}].
\label{eq68}
 \eea

Eq.~(\ref{eq67})  shows a quadratic dependence on the magnetic
field. In the large field regime $C^{(II)}$ contributes as a
constant value to the whole specific heat. Thus, the field
dependence of the specific heat  in the high magnetic fields comes
from $C^{(I)}$ which shows a quadratic behaviour versus the
magnetic field. However, for small field regime we have
$\varepsilon_z\ll\varepsilon_z'$ which makes $C^{II}$ to have the
dominant effect ($C^{(I)}\ll C^{(II)}$). It can be shown that when
the magnetic field is small the leading term of $C^{II}$ is given
by the following expression
\be  C^{(II)}\simeq P'K_B \beta^2
\varepsilon_z'^2\frac{J(J+1)}{3}, \ee
which shows  $B^2$ behaviour for low magnetic fields.

\begin{figure}[t]
\vspace{0.5cm} \centering \scalebox{0.35}[0.35]{
\includegraphics*{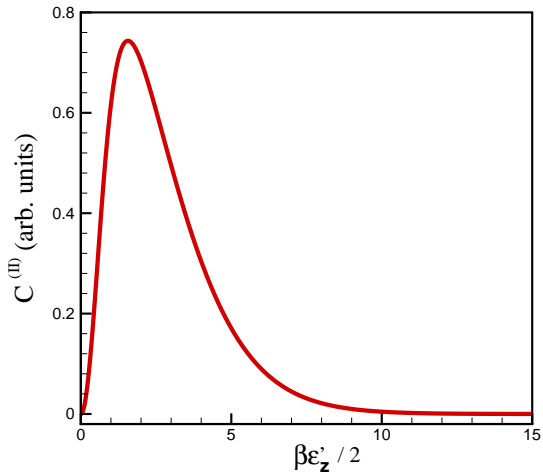}
}
\caption{ Magnetic field (thermal) variation of the $C^{(II)}$.}
\label{fig1}
\end{figure}
%


\subsection{Discussion on the field dependence of the Heat Capacity}

In this section, we have studied the magnetic field dependence of
the Heat capacity of glasses at low temperature. We have assumed
that the tunneling particles carry both an angular momentum and a
nuclear spin. So, the magnetic field couple to both terms.

We have examined our results numerically for arbitrary angle of
EFG and different phases between the quadrupole moments on each
well. The final results are not influenced if we only consider the
simple case of $\theta=0$ and  considering that
$\epsilon_q^R=\epsilon_q^L=\epsilon_q$. Thus,  we have reported
our results in the mentioned special case
 where we can find an analytical expression for the specific heat behaviour.
In this respect,   the total partition function of system can be
splitted  into the
 multiplication of  three separate terms.
 Therefore, the Heat capacity is
composed of three terms, the TLS contribution ($C_{TLS}$), the
nuclear ($C^{(I)}$) and the angular momentum ($C^{(II)}$) parts.
The magnetic field dependence will arise from $C^{(I)}$ and $C^{(II)}$.

We have introduced the generalized Hamiltonian which takes into
account the mentioned degrees of freedom. The direction of
external magnetic ($\theta', \phi'$) field has been considered
arbitrary with respect to the EFG direction of each TLS. However,
the outcome will be averaged over this spherical angle which can
be contracted as a constant to the final results. We have
calculated the contribution of $C_{TLS}$ and $C^{(II)}$ exactly
while the effect of nuclear spin ($C^{(I)}$) has been treated in a
thermal expansion up to the second order corrections.

The different scale of Zeeman energy for the nuclear moment and
the angular momentum of the tunneling entity  is roughly
$\epsilon_z/\epsilon_z'\approx10^{-3}$. Thus, for low magnetic
fields $\beta \epsilon_z'<10$ the dominant effect comes from the
angular momentum part ($C^{(II)}$). Our calculation shows that at
low magnetic field this contribution is proportional to the
square of magnetic field, $C^{(II)}\simeq B^2$ (Eq.(\ref{eq68}).
However, $C^{(II)}$ reaches a maximum at $\beta \epsilon_z'\simeq
3$ and then decreases to zero for $\beta \epsilon_z' >20$ (see
Fig.(\ref{fig1})).

For low magnetic field ($B<10 T$) our result predict a Schottky
like peak in the specific heat of glasses which have nonzero
angular momentum for the tunneling entity. Or at least the sample
has some impurity with nonzero $J$ (angular momentum). This result
is in agreement with the recent experimental observations for
Duran, AlBaSi \cite{strehlow-web} and Suprasil \cite{Str04}. The
data show explicitly an upward increase of the specific heat
versus the magnetic filed for
 $B<0.5 \mbox{T}$ and monotonic decrease
for $B>1 \mbox{T}$ at $T<1 \mbox{K}$.

We have also predicted that for high magnetic fields ($\beta
\epsilon_z' >20$) the dominant contribution to the specific heat
comes from the nuclear moments which shows a quadratic dependence
on the field (Eq.~\ref{eq67}).

In this paper we have studied the special case that the impurities
have only the intrinsic angular momentum part. For the general
case of impurities (for example electrons) with angular momentum,
$J>\frac{1}{2}$ (spin + orbit), we should take into account the
effect of spin - orbit interaction. Which means that at zero
magnetic field the levels $m' = -J,...,J-1,J$  may have different
energies due to the crystal field splitting.

However, if we consider  $J=0$  which means to have a pure glass
without magnetic impurities like BK7 the nuclear quadrupole term
is the only dominant term in the specific heat. So, the  present
calculations predict a quadratic dependence on the magnetic field,
and $1/T^2$ dependence on temperature. As far as we know there is
no experimental result for this case.  Therefore it might be a
good suggestion for future experiments.


\section{Summary}

Based on  the polarization echoes experiments \cite{Nag04} we
believe that the nuclear quadrupole model should
provide the rightful explanation for the intriguing behavior of
the dielectric properties of glasses. A relaxation spectrum rather
sensitive to the orientation of the nuclear quadrupoles is
providing a relaxation susceptibility strongly dependent on the
applied magnetic field \cite{Dor06}, driving us to the conclusion
that the observed magnetic properties of glasses might have a
relaxation origin. Other effects like non-linearities with the
voltage \cite{Bur05}, or cooperative TLS behavior might as well be
found contributing.


\begin{acknowledgments}
A. A. and D. B. would like to acknowledge first of all the
hospitality of the Max Plank Institute for the Physics of Complex
Systems (Dresden, Germany) where this work was carried out, and to
thank Prof. Fulde and Prof W\"{u}rger for stimulating discussions
and valuable comments. A. A. also would like to thank Prof.
Strehlow  for fruitful discussions and useful comments.
\end{acknowledgments}

\begin{appendix}
\section{Generalized TLS Hamiltonian with both nuclear spin and
angular momentum}
The aim of this appendix is to find the generalized TLS
Hamiltonian with both nuclear spin and angular momentum. In the
similar way leading to the Eq.~\ref{eq32}, we can write the total
Hamiltonian as:
\bea
 \left(\begin{array}{cc} H^L+ \frac{\Delta}{2}
\cdot\textbf{1
}&\frac{\Delta_0}{2}\cdot\textbf{1 }\\
\frac{\Delta_0}{2}\cdot\textbf{1 }&H^R- \frac{\Delta}{2}
\cdot\textbf{1 }\end{array}\right) \vspace{0.5cm}
\label{eq69}
\eea
where $\textbf{1}$ is the unit matrix of rank
$(2I+1)\times(2J+1)$. $H^L=V_Q^L+H_Z^L$ and $H^R=V_Q^R+H_Z^R$ are
defined for particles in the left and right wells, respectively.
The state in each well (L or R) can be characterized by $\mid
\psi^{L(R)}_{[I,m],[J,m']}\ket=\mid L(R)\ket\bigotimes \mid
I,m\ket\bigotimes \mid J,m'\ket $ and consequently the matrix
element of the Hamiltonian is defined by the following equation,
\begin{equation}
\label{eq70}
H^{L(R)}_{mm^{\prime},nn^{\prime}}=\bra
\psi^{L(R)}_{[I,m],[J,m']}\mid H^{L(R)}\mid
\psi^{L(R)}_{[I,n],[J,n']}\ket.
\end{equation}
After some simple algebra
we can easily show that for
$H^{R}_{mm',nn'}$ we obtain
\begin{widetext}
\begin{eqnarray} H^{R}_{mm',nn'}&=&\left\{
\epsilon_q^R\left[\Upsilon_{m,m-2}e^{2i\theta}\delta_{m-2,n} +
\Upsilon_{m,m+2}e^{-2i\theta}\delta_{m+2,n}\right]
-\epsilon_z\sin\theta^{'}
\left[e^{i\phi^{'}}\alpha_{-}^m\delta_{m+1,n}+e^{-i\phi^{'}}\alpha_{+}^m\delta_{m-1,n}\right]\right.
\no &+&\left.\left[-m'\epsilon_z'  -\epsilon_z\cos\theta^{'} m +
\epsilon_q^R\Upsilon_{m,m}\right]\delta_{m,n} \right\}
\delta_{m',n'} .
\label{eq71}
\end{eqnarray}
\end{widetext}
where $\theta^{'}$ and $\phi^{'}$ define the direction of magnetic
field in the basis  $(e_1, e_2, e_3)$; $\alpha_{\pm}^m$ and
$\Upsilon_{m,m'}$  are given by Eqs.~(\ref{eq35}~\&~\ref{eq36})  .
 In a similar manner we can obtain the matrix
element of the Hamiltonian for the particles in the left well
$H^{L}_{m,n}$ by taking $\theta=0$ and changing
$\epsilon_q^R\longmapsto\epsilon_q^L$ in the above equation.

\end{appendix}



\end{document}